\documentclass[conference,a4paper]{IEEEtran}
\IEEEoverridecommandlockouts 

\usepackage[cmex10]{amsmath}  
\usepackage{amssymb}
\usepackage{graphicx,color}   

\usepackage{ifxetex}
\ifxetex
\usepackage{xunicode}
\else
\usepackage[utf8]{inputenc}
\usepackage[T1]{fontenc}
\usepackage{microtype}
\fi

\graphicspath{{../pictures/}}

\def\proof{\@IEEElegacywarn{IEEEproof}{IEEEproof}\IEEEproof}

\usepackage{cite}  
\usepackage[ruled,vlined]{algorithm2e}

\newtheorem{theorem}{Theorem}

\newtheorem{lemma}{Lemma}
\newtheorem{definition}{Definition}

\newtheorem{proposition}{Proposition}
\newtheorem{corollary}{Corollary}

\newcommand{\connected}{{N_\textnormal{conn}}}
\newcommand{\remaining}{{N_\textnormal{rem}}}
\newcommand{\stepfour}{{N_\textnormal{(iv)}}}
\newcommand{\tree}{{N_\textnormal{tree}}}
\newcommand{\functionSCC}{{\ttfamily \scshape BreakLeafSCC}}
\renewcommand{\S}{\mathcal{V}_\textnormal{S}} 
\newcommand{\Sp}{\mathcal{V}_\textnormal{S}'}
\newcommand{\Soutside}{\mathcal{V}_\textnormal{S}''}

\def\gap{1.297ex}
  \abovedisplayskip\gap
  \belowdisplayskip\gap
  \abovedisplayshortskip\gap
  \belowdisplayshortskip\gap

\IEEEoverridecommandlockouts
\begin{document}

\title{The Multi-Sender Multicast Index Coding}
\author{\IEEEauthorblockN{Lawrence Ong}
\IEEEauthorblockA{The University of Newcastle,\\ Australia \\
Email: lawrence.ong@cantab.net}
\and
\IEEEauthorblockN{Fabian Lim}
\IEEEauthorblockA{ Massachusetts Institute of Technology,\\ USA\\
Email: flim@mit.edu}
\and
\IEEEauthorblockN{Chin Keong Ho}
\IEEEauthorblockA{Institute for Infocomm Research,\\ Singapore \\
Email: hock@i2r.a-star.edu.sg}
\thanks{This is an extended version of the same-titled paper accepted and to be presented at the IEEE International Symposium on Information Theory (ISIT), Istanbul, in July 2013. This extended paper also contains corrections (highlighted in blue) to missing steps in Algorithm~\ref{algorithm} in the ISIT paper.}
}
\maketitle

\begin{abstract}
We focus on the following instance of an index coding problem, where a set of receivers are required to decode multiple messages, whilst each knows one of the messages a priori.
In particular, here we consider a generalized setting where they are multiple senders, each sender only knows a subset of messages, and all senders are required to collectively transmit the index code.
For a single sender, Ong and Ho (ICC, 2012) have established the optimal index codelength, where the lower bound was obtained using a pruning algorithm. In this paper, the pruning algorithm is simplified, and used in conjunction with an appending technique to give a lower bound to the multi-sender case. An upper bound is derived based on network coding. While the two bounds do not match in general, for the special case where no two senders know any message bit in common, the bounds match, giving the optimal index codelength. The results are derived based on graph theory, and are expressed in terms of strongly connected components.
\end{abstract}

\section{Introduction}

We consider a generalization of the well-known index coding problem to the multi-sender setting where
the senders are \emph{constrained} to know only certain messages. 
As in the typical setup we have $n$ receivers each requiring some of the $m$ independent messages in the set $\mathcal{M} = \{x_1, x_2, \dotsc, x_m\}$, but in addition, we have $S$ separate senders who know only subsets of $\mathcal{M}$.
The problem is precisely given by $(\{\mathcal{M}_s: s \in \{1,2,\dotsc,S\}\}, \{(\mathcal{W}_r, \mathcal{K}_r): r \in \{1,2,\dotsc, n\})$. Here, $(\mathcal{W}_r, \mathcal{K}_r)$ corresponds to receiver $r$, where $\mathcal{K}_r \subseteq \mathcal{M}$ is the subset of messages it knows a priori, and $\mathcal{W}_r \subseteq \mathcal{M}$ is the subset of messages it requires. Furthermore, $\mathcal{M}_s \subseteq \mathcal{M}$ denotes messages that sender $s$ is constrained to know.
Clearly $\mathcal{W}_r \cap \mathcal{K}_r = \{ \}$.
Without loss of generality, we assume  that $\bigcup_{s=1}^S \mathcal{M}_s = \mathcal{M}$, meaning that each message bit is available at some sender(s). 

\begin{figure}[t]
\centering
\includegraphics[width=\linewidth]{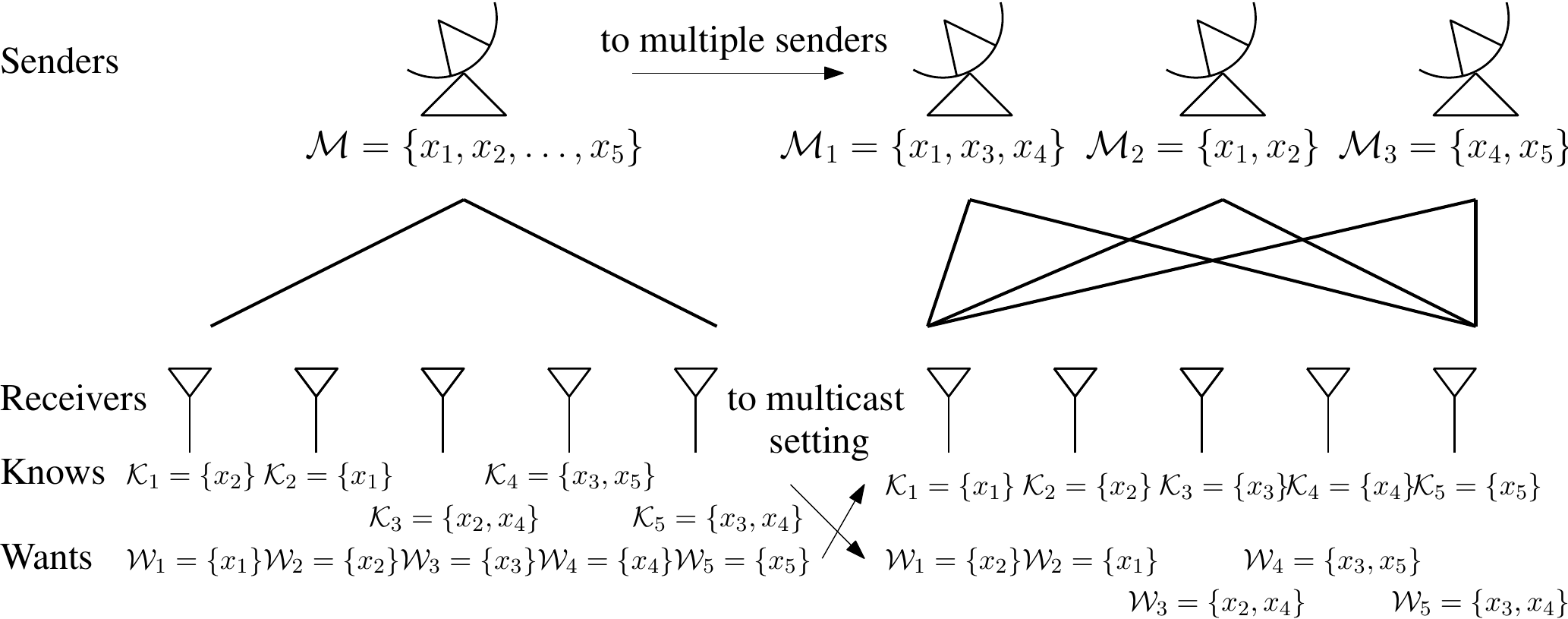}
\caption{Our multi-sender multicast index coding problem. Note two differences with the usual index coding setup. First, we consider distributed-like settings where each sender $s$ is limited to know a subset $\mathcal{M}_s$ of the message set $\mathcal{M}$. Second, we consider the multicast setup where each receiver $r$ knows a unique message, i.e., $\mathcal{K}_r = \{x_r\}$, and wants a larger subset of messages $\mathcal{W}_r$.}
\label{fig-0}
\end{figure}

We define a multi-sender index code for the above setup:
\begin{definition}[Multi-Sender Index Code]
An index code for problem instance $(\{\mathcal{M}_s\}, \{(\mathcal{W}_r, \mathcal{K}_r)\})$ consists of
\begin{enumerate}
\item an encoding function for each sender $s \in \{1,2,\dotsc, S\}$, $E_s: \{0,1\}^{|\mathcal{M}_s|} \mapsto \{0,1\}^{\ell_s}$ such that $\boldsymbol{c}_s = E_s (\mathcal{M}_s)$,
\item a decoding function for each receiver $r \in \{1,2,\dotsc, n\}$, $D_r: \{0,1\}^{\left(\sum_{s=1}^S \ell_s\right)+|\mathcal{K}_r|} \mapsto \{0,1\}^{|\mathcal{W}_r|}$ such that  \\ $\mathcal{W}_r = D_r(\boldsymbol{c}_1, \boldsymbol{c}_2,\dotsc, \boldsymbol{c}_S, \mathcal{K}_r)$.
\end{enumerate}
\end{definition}
That is, each sender encodes its $|\mathcal{M}_s|$-bit message into an $\ell_s$-bit codeword. The codewords of all senders are given to all receivers.
The total number of transmitted bits is thus $\tilde{\ell} = \sum_{s=1}^S \ell_s$, and
$\tilde{\ell}$ the index codelength for the multi-sender generalization of the index coding problem.
See Fig.~\ref{fig-0} for an illustration.
We seek the optimal (i.e., the minimum) index codelength, denoted $\tilde{\ell}^*$, and an optimal index code (i.e., one of length $\tilde{\ell}^*$).
In this paper, we assume that each message bit and each codebit is binary, but our results hold as long as all messages and codeletters take values in the same alphabet.


The case $S=1$ reduces to the usual single-sender index coding problem studied in many works~\cite{birkkol2006,elrouayheb10,baryossefbirk11,dauskachekchee12,ongho12}. This generalization is of interest, for example, in distributed settings where senders are constrained to know only part of the entire message due to, for instance, limited bandwidth between senders for sharing the messages or decoding errors when downloading the messages from a central processor.
Clearly, a multi-sender index code for any $S > 1$ case will also be a code for the single-sender $S=1$ case for the same $\{(\mathcal{W}_r, \mathcal{K}_r)\})$ decoding requirements.
But the converse is not true.
Hence, the techniques described here are new, and previous techniques for the single-sender case do not straightforwardly apply.



This paper also differentiates from other works~\cite{birkkol2006,elrouayheb10,baryossefbirk11,dauskachekchee12} by considering a \emph{multicast} setup.
The classical setup (which is {\em multiprior unicast}) is that each receiver $r$ requires only one unique message (i.e., each $\mathcal{W}_r = \{x_r\}$), knowing a set of message $\mathcal{K}_r$ a priori.
Here we consider the case where each receiver $r$ knows only one unique message (i.e., each $\mathcal{K}_r = \{x_r\}$), but requires a set of messages $\mathcal{W}_r$, which can  be a large subset of $\mathcal{M}$.  We call this the {\em uniprior multicast} problem;
this setup (first looked at in~\cite{ongho12}) is motivated by the {\em multi-way relay channel}~\cite{oechteringschnurr08,ongkellettjohnson12-it}.
The differences between these two setups are depicted in Fig.~\ref{fig-0}.
For the single-sender case, the uniprior multicast problem is completely solved in~\cite{ongho12}, which interestingly shows the optimal scheme to be \emph{linear}; this contrasts with the classical setup where it is known that linear codes can be sub-optimal~\cite{lubertzkystav09}.
Here we explore the multi-sender generalization.
Note that for both unicast and uniprior multicast setups, the number of receivers, $n$, equals the number of messages, $m$.

\section{Graph Representations}

In this paper, we introduce a graphical representation of the uniprior multicast problem $(\{(\mathcal{M}_s)\},\{(\mathcal{W}_r,x_r)\})$ to capture both decoding requirements and sender constraints. This graphical representation shall be useful for stating and proving our subsequent results.

A graph $\mathcal{G}= (\mathcal{V},\mathcal{A})$ is a tuple of a vertex set $\mathcal{V}$ and an arc/edge set $\mathcal{A}$.
We correspond a vertex to each of the $n$ receivers.
An arc $(i \rightarrow j)$ conveys directional information from vertex $i$ to vertex $j$, while an edge $(i,j)$ is undirected.
An (undirected) graph has only edges, while a (directed) \emph{digraph} has only arcs; both cannot have self-loops.
We represent the multi-sender uniprior multicast problem as follows.


\begin{definition}
The decoding requirements determined from $\{(\mathcal{W}_r,x_r)\}$ is represented by an $n$-vertex {\em information-flow} digraph\footnote{The definition of the information-flow graph defined for the uniprior multicast problem here (where the arcs capture what the receivers {\em require}) is different from the side-information graph  defined for the unicast problem~\cite{baryossefbirk11} (where the arcs capture what the receivers {\em know}).} $\mathcal{G}= (\mathcal{V},\mathcal{A})$, where $(i \rightarrow j) \in \mathcal{A}$ if and only if (iff) receiver $j$ requires $x_i$, i.e., $x_i \in \mathcal{W}_j$.
The sender constraints determined from $\{\mathcal{M}_s\}$ is represented by an $n$-vertex \emph{message} graph $\mathcal{U} = (\mathcal{V}, \mathcal{E})$, where $(i, j) \in \mathcal{E}$ iff messages $x_i$ and $x_j$ are known to the same sender, i.e., $i,j \in \mathcal{M}_s$ for some $s$. Note, both $\mathcal{G}$ and $\mathcal{U}$ share the \emph{same} vertex set $\mathcal{V}$.
\end{definition}

We denote the optimal index codelength for the problem represented by $(\mathcal{G},\mathcal{U})$ as $\tilde{\ell}^*(\mathcal{G},\mathcal{U})$.

In the sequel, we work on the {\em simplified model} described as follows: For any index coding problem $(\mathcal{G}_1, \mathcal{U}_1)$, we construct a {\em simplified} $(\mathcal{G}_2, \mathcal{U}_2)$ by removing every message $x_i$  that is not required by any receiver (meaning that vertex $i$ has no outgoing arc in $\mathcal{G}_1$) from the receivers and the senders; equivalently, we set $x_i = \varnothing$. There is no loss of generality because an optimal index code for $(\mathcal{G}_2, \mathcal{U}_2)$ is an {\em optimal} index code for $(\mathcal{G}_1,\mathcal{U}_1)$, see proof in Appendix~A.

%

Note that an edge $(i,j)$ in the message graph does not indicate which sender(s) owns both the messages (i.e. the set of $s$ such that $i,j \in\mathcal{M}_s$).
This ambiguity will not affect the techniques developed in this paper, though we will point out in the conclusion the existence of certain multi-sender uniprior multicast problems where
resolving this ambiguity may lead to further improvement of our results.

\subsection{Terminology}

We will use common graph terminology~\cite{Bang-JensenGutin}:
A {\em strongly connected component} (SCC) of a digraph is a {\em maximal} subgraph of the digraph such that in the subgraph, for any vertex pair $i,j$, there is a directed path from $i$ to $j$ and another from $j$ to $i$.
A {\em leaf} vertex in a digraph has no outgoing arcs.
A vertex $j$ is a {\em predecessor} of vertex $i$ iff there is a directed path from $j$ to $i$. A {\em tree} is a connected undirected subgraph with no cycle.


\section{Results} \label{section:results}


The main contribution of this paper is the technique we propose to obtain a lower bound to $\tilde{\ell}^*$, which will be tight in a few cases.
The lower bound and achievability in this paper will be stated in terms of {\em leaf SCCs} in the information-flow digraph $\mathcal{G}$. A leaf SCC is an SCC that has (a) no outgoing arc (i.e., from a vertex in the SCC to a vertex outside the SCC), and (b) at least two vertices (i.e., {\em non-trivial}).

\subsection{A New Technique for the Single-Sender Problem}

We first obtain results for the single-sender case (i.e., with no sender constraint). While the single-sender case has been solved~\cite{ongho12}, we propose an alternative technique to obtain a lower bound to the index codelength. This then allows us to develop intuition on the main results for the general multi-sender case.
This intuition is based on the following lemmas:

\begin{lemma} \label{lemma:predecessor}
From any index code, each receiver $i$ must be able to decode the messages of all its predecessors. 
\end{lemma}

\begin{lemma} \label{lemma:predecessor-all}
From any index code, any receiver must be able to decode the messages of all predecessors of any leaf vertex.
\end{lemma}

Lemma \ref{lemma:predecessor-all}  (proven in Appendix~B) is a simple corollary to Lemma \ref{lemma:predecessor}. 
The proof of former is as follows:
\begin{IEEEproof}[Proof of Lemma \ref{lemma:predecessor}]
For every arc $(j \rightarrow i)$, receiver $i$ must be able to decode $x_j$. Having decoded $x_j$, receiver $i$ knows the only a priori message that receiver $j$ has. Therefore, receiver $i$ must also be able to decode all messages required (and hence decodable) by receiver $j$, i.e.,
$\mathcal{W}_j = \{x_k: (k \rightarrow j) \in \mathcal{A}\}$. Further chaining of this argument shows that receiver $i$ must be able to decode messages $x_j$ of all  predecessors $j$ of $i$. 
\end{IEEEproof}


We now use the above lemmas to develop a simple lower bound to the optimal index codelength. 
First, we make some useful graph-related definitions.
A leaf SCC with vertex set $\S$ is said to be {\em pruned}, if one vertex $v \in \S$ is arbitrarily selected, and all outgoing arcs are removed from $v$.
A vertex is said to be {\em grounded} if the vertex is a leaf vertex or a predecessor of some leaf vertex.
A digraph is said to be grounded if every vertex in the digraph is grounded.
For example, directed tree is grounded; more generally, a digraph that contains no leaf SCC is grounded (see Appendix~C). 
 
Consider $(\mathcal{G},\mathcal{U})$.
We prune all leaf SCCs in $\mathcal{G}$ to obtain a grounded digraph $\mathcal{G}^\dagger$. 
Let $V_\textnormal{out}(\mathcal{G^\dagger})$ denote the number of non-leaf vertices in $\mathcal{G^\dagger}$ (i.e., each with at least one outgoing arc).
Then invoking Lemma~\ref{lemma:predecessor-all},
any receiver (even those with no prior, i.e., $\mathcal{K}_j= \{ \}$) must be able to decode
$V_\textnormal{out}(\mathcal{G}^\dagger)$ message bits (the messages of all non-leaf vertices in $\mathcal{G}^\dagger$).
That is, we must have the lower bound $\tilde{\ell}^*(\mathcal{G}^\dagger, \mathcal{U}) \geq V_\textnormal{out}(\mathcal{G}^\dagger)$. Note that pruning a leaf SCC reduces decoding requirements, and hence $\tilde{\ell}^*(\mathcal{G},\mathcal{U}) \geq \tilde{\ell}^*(\mathcal{G}^\dagger,\mathcal{U})$. Consequently, we have $\tilde{\ell}^*(\mathcal{G},\mathcal{U}) \geq V_\textnormal{out}(\mathcal{G}^\dagger)$. This lower bound---which does not take $\mathcal{U}$ into account---is tight for the single-sender case.\footnote{Due to different approaches, the lower bound here takes a form different from \cite{ongho12}. However, one can show that they are equivalent.}

The lower bound $V_\textnormal{out}(\mathcal{G}^\dagger)$ derived above applies to any $\mathcal{U}$ (i.e., both single-sender and multi-sender) as long as $\mathcal{G}$ remains the same.
Unfortunately, in our multi-sender setting, this lower bound may be \emph{unachievable} due to the sender constraints (captured by the message graph $\mathcal{U})$.
There are specific situations where this lower bound is achievable. 
Consider the case where every leaf SCC of $\mathcal{G}$ has the property that, there always exists a path in $\mathcal{U}$ between any two vertices in the SCC, where this path is constrained to involve only vertices within the SCC.
We say that such leaf SCCs are {\em message-connected}.
If all leaf SCCs in $\mathcal{G}$ are message-connected, then a similar scheme to the single-sender case~\cite{ongho12} achieves $V_\textnormal{out}(\mathcal{G}^\dagger)$ bits. 
We clarify the schemes in the later Sec.~\ref{section:achievable}.


\subsection{A Tighter Lower Bound to $\tilde{\ell}^*$}
The above-mentioned lower bound can be tightened by considering the sender constraints. To this end, we introduce new techniques in this paper, which requires the following further characterization of the leaf SCCs in $\mathcal{G}$ in relation to the message graph $\mathcal{U}$. 
\begin{enumerate}
\item Recall that a leaf SCC is {\em message-connected} iff a path exists in $\mathcal{U}$ between any two vertices in the SCC, and this path consists of vertices within the SCC.
\item A leaf SCC is {\em message disconnected} iff there are two vertices in the SCC with no path in $\mathcal{U}$ between them;
\item A leaf SCC which is neither message connected nor message disconnected is {\em semi message connected}, referred also as semi leaf SCC. Thus, there exists a path in $\mathcal{U}$ between two vertices in the SCC with at least one vertex of the path outside the SCC.
%
%
\end{enumerate}

Semi leaf SCCs are also further classified using the following property.
For a vertex set $\S\subseteq\mathcal{V}$, a vertex $i \notin \S$ is an \emph{m-neighbor} of $\S$ iff there is an edge $(i,v)$ between $i$ and some $v \in \S$.
Here, ``m'' is mnemonic for message.
Then, a semi leaf SCC with vertex set $\S$, is said to be {\em degenerated} iff
\begin{enumerate}
\item $\S$ can be partitioned into \emph{two} parts $\Sp$ and $\S\setminus \Sp$ such that there is no edge in $\mathcal{U}$ across vertices from different parts, and
\item there exists a vertex subset not in $\S$, denoted by $\Soutside \subseteq \mathcal{V} \setminus \S$, which
\begin{itemize}
\item can only have at most one non-leaf vertex (the other vertices must strictly be leaf)
\end{itemize}
such that
\begin{itemize}
\item every m-neighbor of $\Sp$ is in $\Soutside$ or is a predecessor of some vertex in $\Soutside$.
\end{itemize}
\end{enumerate}

\begin{figure}[t]
\centering
\includegraphics[width=\linewidth]{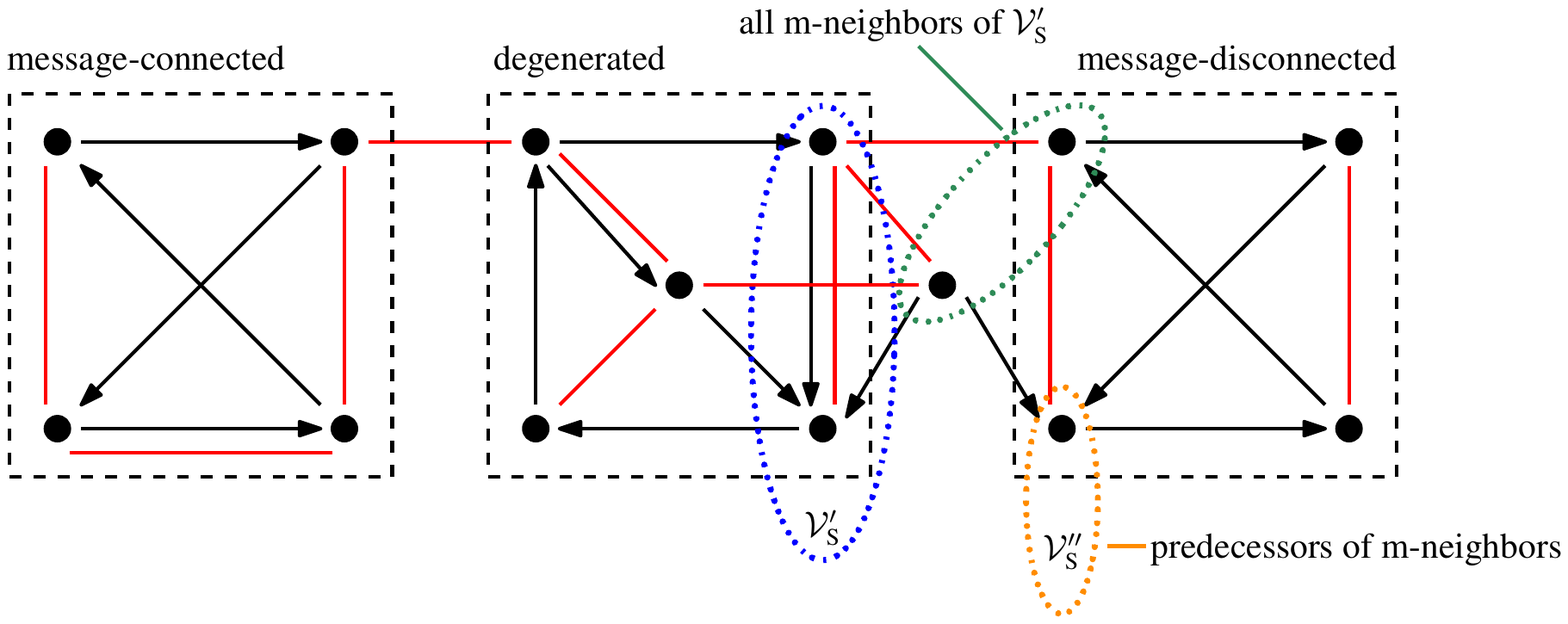}
\caption{Example of an index coding problem, diagrammed by super-imposing an information-flow digraph $\mathcal{G}$ (arcs in black) and a message graph $\mathcal{U}$ (edges in red).
By their definition, leaf SCCs are determined by $\mathcal{G}$, but their various types are determined also in accordance with $\mathcal{U}$.
These graphs illustrate concurrently three leaf SCCs types: (i) message connected, where there is a (red) path between any two vertices through only vertices in the SCC; (ii) message disconnected, containing two vertices cannot be connected with a red path; and (iii) semi message connected, where some vertices must be connected by a path with vertices outside the SCC. Note, the semi leaf SCC here is degenerated because we can find two vertex sets $\mathcal{V}_\text{S}'$ and $\mathcal{V}_\text{S}''$, such that all m-neighbors of $\mathcal{V}_\text{S}'$  are predecessors of $\mathcal{V}''_\text{S}$. }
\label{fig-1}
\end{figure}





Fig.~\ref{fig-1} illustrates the above characterizations of leaf SCCs.
The discussion in the previous section explains that a grounded graph $\mathcal{G}^\dagger$ delivers a lower bound to $\tilde{\ell}^*(\mathcal{G},\mathcal{U})$.
We will describe new techniques to arrive at ``better'' grounded graphs $\mathcal{G}^\dagger$ from the original digraph $\mathcal{G}$, giving tighter lower bounds
\begin{equation}
\tilde{\ell}^*(\mathcal{G}^\dagger,\mathcal{U}^\dagger) \geq V_\textnormal{out}(\mathcal{G}^\dagger),  \label{eq:lower}
\end{equation}
where $\mathcal{U}^\dagger$ is the message graph that may be modified from the original $\mathcal{U}$. Here, we do not prune all leaf SCCs (pruning reduces $V_\textnormal{out}(\mathcal{G}^\dagger)$), and this gives us a larger $V_\textnormal{out}(\mathcal{G}^\dagger)$.

If a digraph contains no leaf SCC, then it is grounded (proven in Appendix~C). So, to produce the required $(\mathcal{G}^\dagger,\mathcal{U}^\dagger)$, we devise Algorithm~\ref{algorithm} given below.
This algorithm iteratively ``breaks'' leaf SCCs or changes arc-connectivity of the vertices therein, such that the resultant $\mathcal{G}^\dagger$ contains no leaf SCC.
The algorithm is also specially designed to make sure that in each step, the optimal index codelength cannot increase, thus giving
\begin{equation}
\tilde{\ell}^*(\mathcal{G},\mathcal{U}) \geq \tilde{\ell}^*(\mathcal{G}^\dagger,\mathcal{U}^\dagger). \label{eq:grounded}
\end{equation}
Combining \eqref{eq:lower} and \eqref{eq:grounded}, we have the lower bound
\begin{equation}
\tilde{\ell}^*(\mathcal{G},\mathcal{U}) \geq V_\textnormal{out}(\mathcal{G}^\dagger). \label{eq:lower-bound-1}
\end{equation}


\begin{algorithm}[t]
\begin{small}
\SetKwBlock{Begina}{function {\normalfont \functionSCC}}{end}
\Begina{
\ForEach{message-connected leaf SCC}{
(i) Prune the leaf SCC\;
}
\While{\color{blue} there exists message-disconnected or degenerated SCC}{
\ForEach{message-disconnected leaf SCC}{
(ii) Append a dummy vertex to the leaf SCC\;
}
\While{there exists degenerated leaf SCC, with vertex set $\S$,}{
\If{the set $\Soutside \subseteq \mathcal{V}\setminus \S$ contains one non-leaf vertex, denoted by $v'' \in \Soutside$,}{
(iii-a) Add an arc from any vertex in $\Sp$ to the vertex $v''$\;}
\Else{(iii-b) Add an arc from any vertex in $\Sp$ to any vertex in $\Soutside$\;}
}
}
}
\BlankLine
\Begin(\tcp*[f]{The algorithm starts here}){
\tcp{Phase 1}
Run {\functionSCC}\;
\BlankLine
\tcp{Phase 2: Iteration \& optimization}
\While{there exists leaf SCC}{
\If{\color{blue} there exists message-connected leaf SCC}{
 \color{blue} (iv-0) Run {\functionSCC} but execute step (i) only once (i.e., prune only one message-connected leaf SCC)\;}
\Else{
(iv-a) Select one semi leaf SCC\;
(iv-b) Arbitrarily add edges between vertex pairs until the leaf SCC is message connected\;
(iv-c) Run {\functionSCC}\;
}
}
}
\caption{Breaking all Leaf SCCs}
\label{algorithm}
\end{small}
\end{algorithm}


We propose different methods to {\em break} the leaf SCCs in $\mathcal{G}$, such that they are no longer leaf SCCs.
Algorithm~\ref{algorithm} has two distinct phases.
In phase 1, the procedure {\functionSCC} is run once to break all initial message-connected leaf SCCs  in $\mathcal{G}$, and then all message-disconnected and degenerated leaf SCCs.
Note that some leaf SCCs may be made message-connected or message-disconnected in step (iii). Also, pruning message-connected leaf SCCs might make some semi leaf SCCs degenerated. Phase 2 breaks the remaining leaf SCCs. 

Step (ii) in {\functionSCC} involves \emph{appending} a dummy leaf vertex $i$ to a leaf SCC (with vertex set $\S$), by appending an outgoing arc $(v\rightarrow i)$ from some $v \in \S$ to $i$.
In step (iv-a), the choice of leaf SCC for adding edges is arbitrary. Nevertheless, a proper choice of leaf SCC(s) \emph{minimizes} the number of rounds that step (iv) iterates, which we shall see will then give a better lower bound.

Let $\connected$ denote the number of message-connected leaf SCCs in $\mathcal{G}$. Note that $\connected$ for each $\mathcal{G}$ is fixed, independent of the algorithm.
Let $\remaining$ denote the number of leaf SCCs (they must be semi leaf SCCs) that remain after the initial run of {\functionSCC} in phase 1, and $\stepfour$ denote the number of iterations in phase 2.
We will show (in Sec.~\ref{sec:remove-scc}) that step (iv) always reduces the number of leaf SCCs, and so $\stepfour \leq \remaining$.
We now state the main result of this paper:
\begin{theorem} [Lower bound] \label{theorem:multi-sender-lower-bound}
The optimal multi-sender index codelength is lower bounded as
\begin{equation}
\tilde{\ell}^* \geq V_\textnormal{out}(\mathcal{G}) - ( \connected + \stepfour). \label{eq:lower-bound-2}
\end{equation}
\end{theorem}


We will prove Thm.~\ref{theorem:multi-sender-lower-bound} in Sec.~\ref{section:upper}. We will see that the RHS of \eqref{eq:lower-bound-2} equals the RHS of \eqref{eq:lower-bound-1}. As mentioned earlier, the lower bound is optimized by finding the smallest $\stepfour$.

\subsection{Achievability}

Our achievability scheme is based on the construction of special trees  in the message graph $\mathcal{U}$, referred to as {\em connecting trees}, which
has all the following properties placed on its vertex set $\mathcal{V}^\textnormal{T}$:
\begin{enumerate}
\item Each vertex in $\mathcal{V}^\textnormal{T}$ has one or more outgoing arcs in $\mathcal{G}$.
\item Each vertex in $\mathcal{V}^\textnormal{T}$ has no outgoing arc in $\mathcal{G}$ to $\mathcal{V} \setminus \mathcal{V}^\textnormal{T}$,


\item No vertex in $\mathcal{V}^\textnormal{T}$ belongs to any message-connected leaf SCCs or another connecting tree.
\end{enumerate}
Let $\tree$ denote the number of connecting trees that can be found. We will propose a coding scheme that achieves the following index codelength:

\begin{theorem}[Achievability] \label{theorem:multi-sender-achievable}
The optimal multi-sender index codelength is upper bounded as
\begin{equation}
\tilde{\ell}^* \leq V_\textnormal{out}(\mathcal{G}) - ( \connected + \tree).
\end{equation}
\end{theorem}

We prove Thm.~\ref{theorem:multi-sender-achievable} in Sec.~\ref{section:achievable}. The achievability is optimized by finding the maximum number of connecting trees.

\begin{figure}[t]
\centering
\includegraphics[width=4cm]{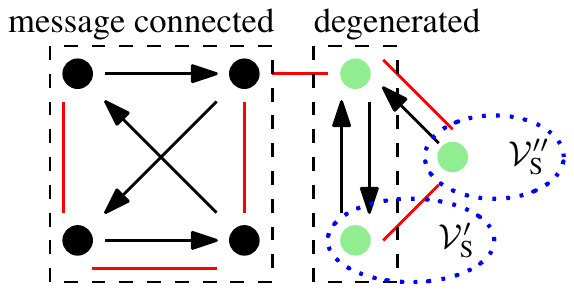}
\caption{An index coding problem represented by $\mathcal{G}$ (with black arcs) and $\mathcal{U}$ (with red edges). There are two leaf SCCs in the graph: (i) message connected, and (ii) degenerated. Here, we can form a connecting tree using the green vertices. }
\label{fig-2}
\end{figure}

For Fig.~\ref{fig-1}, we cannot form any connecting tree; for Fig.~\ref{fig-2}, we can form one connecting tree using the green vertices.

\subsection{Special Cases}

Combining Thms.~\ref{theorem:multi-sender-lower-bound} and \ref{theorem:multi-sender-achievable}, we conclude $\stepfour \geq \tree$, and thus the optimal index codelength is found within $\stepfour - \tree$ bits.
In the following special cases, we have $\stepfour = \tree$, and the lower bound is tight.




\begin{corollary} \label{corollary:no-semi}
If no leaf SCC remains after running phase 1 of Algorithm~\ref{algorithm}, then
\begin{equation}
\tilde{\ell}^* = V_\textnormal{out}(\mathcal{G}) - \connected. \label{eq:unique}
\end{equation}
\end{corollary}

\begin{IEEEproof}
Since $\remaining \geq \stepfour \geq \tree$, $\remaining = 0$ implies that $\stepfour = \tree = 0$.
\end{IEEEproof}

\begin{corollary} \label{corollary:unique}
If each bit $x_i$ in the message set $\mathcal{M}$ is known to only one sender (i.e., the $n$ sender constraint sets $\mathcal{M}_s$ partition $\mathcal{M}$),
then the optimal index codelength is given by \eqref{eq:unique}.
\end{corollary}

\begin{IEEEproof}
If messages $x_i$ and $x_j$ belong to some sender $s$ (i.e. $x_i, x_j \in \mathcal{M}_s$), then there exists an edge $(i,j)$ in the message graph $\mathcal{U}$.
Otherwise, if the messages $x_i,x_j$ belong to different senders, it is impossible to have a {\em path} between $i$ and $j$.
This means we have only message connected or disconnected leaf SCCs, i.e.,
there is no semi leaf SCC. Thus, $\remaining=0$.
\end{IEEEproof}

Corollary~\ref{corollary:unique} includes the result of the single-sender problem~\cite{ongho12} as a special case.

\begin{figure}[t]
\centering
\includegraphics[width=3.5cm]{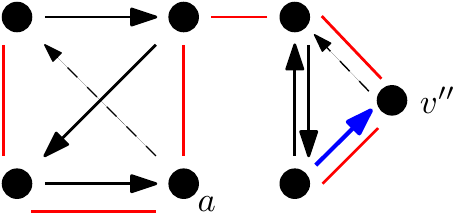}
\caption{The resultant graph after running Algorithm~\ref{algorithm} on Fig.~\ref{fig-2}, where the dashed arcs are removed, and the blue arc added. }
\label{fig-3}
\end{figure}

\subsection{An Example}

We now illustrate the lower bound and the achievability for the graph in Fig.~\ref{fig-2}. For lower bound, we run Algorithm~\ref{algorithm} to obtain the resultant graph shown in Fig.~\ref{fig-3} (with dashed arcs removed and blue edge/arc added). Specifically, starting from Fig.~\ref{fig-2}, note that the furthest-left four vertices form a message-connected leaf SCC. Executing step (i) of the algorithm, we remove all outdoing arcs from a vertex. Here, vertex $a$ (in Fig.~\ref{fig-3}) is arbitrarily chosen, and the dashed arc from $a$ is removed. The right three vertices form a degenerated leaf SCC. Executing step (iii-a), we add an arc (indicated by the blue arc in Fig.~\ref{fig-3}) from the vertex in $\Sp$ to  $v'' \in \Soutside$. Entering phase 2, we note that the right three vertices now form a message-connected leaf SCC. Executing step (iv-0), we prune the leaf SCC by removing the dashed arc. The Algorithm terminates here, after breaking all leaf SCCs. From Theorem~\ref{theorem:multi-sender-lower-bound}, we have the lower bound $\tilde{\ell}^* \geq 7 - (1 + 1) = 5$.

For achievability, recall that we can form a connecting tree using the three green vertices in Fig.~\ref{fig-2}. So, Theorem~\ref{theorem:multi-sender-achievable} gives the upper bound $\tilde{\ell}^* \leq 7 - (1 + 1) = 5$.

For this example, the optimal index codelength is 5 bits.

\section{Proof of Theorem~\ref{theorem:multi-sender-lower-bound} (Lower Bound)} \label{section:upper}
We will refer to each vertex $i$ as receiver $i$ (and vice versa), and $x_i$ as the message of receiver/vertex $i$.




In Sec.~\ref{sec:remove-scc}, we will prove that Algorithm~\ref{algorithm} produces a grounded $\mathcal{G}^\dagger$, meaning that \eqref{eq:lower} holds; in Sec~\ref{sec:cannot-increase-length}, we will prove that $\mathcal{G}^\dagger$ produced by Algorithm~\ref{algorithm} satisfies \eqref{eq:grounded}.


\subsection{Algorithm~\ref{algorithm} Produces a Grounded Digraph} \label{sec:remove-scc}

We now show that Algorithm~\ref{algorithm} produces a digraph with no leaf SCC, and hence grounded. As the algorithm terminates after all leaf SCCs have been broken, it suffices to show that the algorithm always terminates, i.e., step (iv) iterates for a finite number of times. This is true if step (iv) always reduces the number of leaf SCCs.  

We first show that any of the steps (i), (ii), and (iii-b) reduces the number of leaf SCCs by one. In step (i), after removing all outgoing arcs from some vertex $v$ in a leaf SCC, $v$ and other vertices in the SCC (each having a directed path to $v$) are grounded. In steps (ii) and (iii-b), an arc is added from a leaf SCC to a leaf vertex. This will also ground all vertices in the SCC. As any grounded vertex cannot belong to a leaf SCC, each of these steps breaks the leaf SCC it ``operates'' on, thereby reducing the number of leaf SCCs by one.

We now show that step (iii-a) cannot increase the number of leaf SCCs.
Step (iii-a) adds an arc from a leaf SCC (denote by $\mathcal{G}_\text{leaf}$) to some vertex $v''$ not in $\mathcal{G}_\text{leaf}$. There are three possibilities: (1) $v''$ is grounded. Using the argument for step (ii), the number of leaf SCCs decreases by one. (2) $v''$ is not grounded and has no directed path to $\mathcal{G}_\text{leaf}$. In this case $\mathcal{G}_\text{leaf}$ is made non-leaf, and the number of leaf SCCs decreases by one. (3) $v''$ is not grounded and has a directed path to $\mathcal{G}_\text{leaf}$. In this case the SCC $\mathcal{G}_\text{leaf}$ expands to include more vertices (including $v''$) and arcs. The number of leaf SCCs  decreases (if the expanded SCC is non-leaf) or stays the same (otherwise).







Finally, consider each iteration (iv). Step (iv-b) makes a semi leaf SCC message connected.  This step, only adding edges, does not change the number of leaf SCCs. When running {\functionSCC} in step (iv-c), the leaf SCC that has been made message-connected will be broken in step (i), and other steps (ii) and (iii) cannot increase the number of leaf SCCs. So, step (iv) always reduces the number of leaf SCCs.

\subsection{Algorithm~\ref{algorithm} Cannot Increase the Optimal Codelength} \label{sec:cannot-increase-length}

Now, we prove \eqref{eq:grounded} by showing that each of the steps (i)--(iv) cannot increase the optimal index codelength, i.e.,
\begin{equation}
\tilde{\ell}^* ( \mathcal{G}', \mathcal{U}') \leq \tilde{\ell}^*(\mathcal{G},\mathcal{U}), \label{eq:g-prime-ell}
\end{equation}
where  $\mathcal{G} = (\mathcal{V}, \mathcal{A})$ and $\mathcal{U} = (\mathcal{V}, \mathcal{E})$ respectively denote
information-flow and message graphs before each of the steps (i)--(iv); $\mathcal{G}' = (\mathcal{V}', \mathcal{A}')$ and $\mathcal{U}' = (\mathcal{V}', \mathcal{E}')$, after the step.

\subsubsection{Step (i)}
Removing arcs in the information-flow digraph is equivalent to removing decoding requirements for the receivers. Hence, we have \eqref{eq:g-prime-ell}.

\subsubsection{Step (ii)}
Now, as adding an arc $(v \rightarrow i)$ and a dummy receiver $i$ increases decoding requirements, we have $\tilde{\ell}^* ( \mathcal{G}', \mathcal{U}') \geq \tilde{\ell}^*(\mathcal{G},\mathcal{U})$.
But, Lemma~\ref{lemma:disconnected} below says that using any optimal index code for $(\mathcal{G},\mathcal{U})$---of length $\tilde{\ell}^*(\mathcal{G},\mathcal{U})$---a dummy receiver can decode all messages of all message-disconnected leaf SCCs.
Hence, this index code also satisfies the decoding requirements of $(\mathcal{G}',\mathcal{U}')$, i.e., $\tilde{\ell}^*(\mathcal{G},\mathcal{U})$ is achievable for the problem instance $(\mathcal{G}',\mathcal{U}')$. So, \eqref{eq:g-prime-ell} in fact holds with equality.

\begin{lemma} \label{lemma:disconnected}
For any index code, any receiver is able to decode the messages of all message-disconnected leaf SCCs.
\end{lemma}

\begin{IEEEproof}
Let $\S$ be the vertex set of a message-disconnected leaf SCC.
By definition, we can partition all vertices $\mathcal{V}$ into two non-empty sets $\mathcal{V}_1$ and $\mathcal{V}_2$, such that two vertices $a,b \in \S$ in the leaf SCC lie on separate partitions and  cannot have an undirected path (in $\mathcal{U}$) between them. Let
$a \in \mathcal{V}_1$ and $b \in \mathcal{V}_2$.

The lack of edge-connectivity between $\mathcal{V}_1$ and $\mathcal{V}_2$ implies that any index code can be partitioned into two parts, $\boldsymbol{c} = (\boldsymbol{c}_1, \boldsymbol{c}_2)$, such that
every bit in $\boldsymbol{c}_1$  depends on only $\{x_i: i \in \mathcal{V}_1\}$ and not on $\{x_j: j \in \mathcal{V}_2\}$, and vice versa.

Since $a$ and $b$ belong to an SCC,
receiver $a$ must decode $x_b$ (see Lemma~\ref{lemma:predecessor}). Note that $\boldsymbol{c}_1$ (which contains $x_a$) and $\boldsymbol{c}_2$ do not have any message bit in common. So, knowing $x_a$ can help receiver $a$ decode only messages in $\boldsymbol{c}_1$.
Also, receiver $a$ must decode $x_b$ using solely $\boldsymbol{c}_2$, without using its prior. 
Hence, if $a$ can decode $x_b$, so can any receiver---even one without prior.
Since the choice of $a,b$ was arbitrary, we have Lemma~\ref{lemma:disconnected}.
%
%
\end{IEEEproof}


\subsubsection{Step (iii)}

Recall the vertex subsets $\Sp \subset \S$ and $\S''\subseteq \mathcal{V} \setminus \S$. In step (iii), we append an arc from $\Sp$ to some vertex $v'' \in \S''$. 
If we can show that receiver $v''$ can decode all messages of $\Sp$ (using any index code for $(\mathcal{G},\mathcal{U})$), then by the arguments for step (ii), we conclude \eqref{eq:g-prime-ell} holds with an equality.

By Lemmas~\ref{lemma:predecessor} and \ref{lemma:predecessor-all}, receiver $v''$ can decode the messages of all m-neighbors of $\Sp$, denoted as $\mathcal{N}(\Sp)$, as each vertex in $\mathcal{N}(\Sp)$ is either grounded, a predecessor of $v''$, or  $v''$ itself.


Let $a \in \S \setminus \Sp$ and $b \in \Sp$, where $a$ must decode $x_b$. We will show that if $a$ can decode $x_b$, so can $v''$. Consequently, $v''$ can decode all messages of $\Sp$.
There is no edge across $\Sp$ and $\S \setminus \Sp$, meaning that any index codebit cannot be a function of messages from both the sets. So, we can partition any index code for $(\mathcal{G},\mathcal{U})$ into $\boldsymbol{c} = (\boldsymbol{c}_1, \boldsymbol{c}_2)$, where
$\boldsymbol{c}_1$ does not contain  any message of $\Sp$, and
$\boldsymbol{c}_2$ contains  only the messages of $\Sp$ and $\mathcal{N}(\Sp)$.
Any advantage in decoding that $a$ has over $v''$ is due to knowing $x_a$, but $x_a$ can help $a$ only in decoding the messages in $\boldsymbol{c}_1$, which $a$ can then use to decode the messages in $\boldsymbol{c}_2$ (which contain $x_b$).
As $v''$ is able decode the messages of $\mathcal{N}(\Sp)$, which contains all the overlap of messages in $\boldsymbol{c}_1$ and $\boldsymbol{c}_2$, $v''$  is as capable as $a$ in decoding $x_b$.

\subsubsection{Step (iv-b)}
Appending an edge $(i,j)$ is equivalent to appending messages $\{x_i,x_j\}$ to some sender's message set $\mathcal{M}_s$.
Doing so relaxes the sender constraints, and thus the optimal codelength can only decrease. Hence \eqref{eq:g-prime-ell} holds.

\subsection{Evaluating the Lower Bound}


Combining \eqref{eq:lower} and \eqref{eq:grounded}, we have \eqref{eq:lower-bound-1}. We now show \eqref{eq:lower-bound-2}.

{\color{blue}In each iteration (iv), exactly one message-connected leaf SCC is pruned, either in step (iv-0) or in  step (iv-c), which executes step (i) {\em once} through {\functionSCC}.
Note that step (iv-b) makes only {\em one} leaf SCC  message connected. 
Out of all the steps, only pruning, i.e., step (i), changes $V_\textnormal{out}(\cdot)$---reducing it by one at each run.} So when  Algorithm~\ref{algorithm} terminates, we have
\begin{align}
V_\textnormal{out}(\mathcal{G}^\dagger) & = {\color{blue} V_\textnormal{out}(\mathcal{G}) - N_\textnormal{(i) in phase 1} - N_\textnormal{(i) in phase 2}} \nonumber \\
& = {\color{blue} V_\textnormal{out}(\mathcal{G}) - N_\textnormal{(i) in phase 1} - N_\textnormal{(iv-0)} - N_\textnormal{(iv-c)}} \nonumber \\
& = V_\textnormal{out}(\mathcal{G}) - (\connected + N_\textnormal{(iv)}), \label{eq:count-steps}
\end{align}
\noindent where $N_\textnormal{($a$)}$ is the number of times step ($a$) is run. 
From \eqref{eq:lower-bound-1} and \eqref{eq:count-steps}, we have \eqref{eq:lower-bound-2}. This completes the proof of Thm.~\ref{theorem:multi-sender-lower-bound}. $\hfill\blacksquare$





\section{Proof of Theorem~\ref{theorem:multi-sender-achievable} (Achievability)} \label{section:achievable}

In this section, we will show that there exists an index code of length $\tilde{\ell} = V_\textnormal{out}(\mathcal{G}) - (\connected + \tree)$. 
Let a set of connecting trees be $\{\mathcal{T}_t = (\mathcal{V}^\textnormal{T}_t, \mathcal{E}^\textnormal{T}_t): t \in \{1,2,\dotsc, \tree\} \}$, and all the message-connected leaf SCCs in $\mathcal{G}$ be $\{\mathcal{C}_c = (\mathcal{V}^\textnormal{C}_c, \mathcal{A}^\textnormal{C}_c): c \in \{1,2,\dotsc, \connected\}\}$. Further, let the remaining vertices in $\mathcal{G}$ be $\mathcal{V}' = \mathcal{V} \setminus \{ \bigcup_{t=1}^\tree \mathcal{V}^\textnormal{T}_t \cup \bigcup_{c=1}^\connected \mathcal{V}^\textnormal{C}_c \}$. Denote by $\mathcal{V}'_\textnormal{out}$ the set of all non-leaf vertices in $\mathcal{V}'$. By definition, all $\mathcal{V}^\textnormal{T}_t,\mathcal{V}^\textnormal{C}_c,$ and $\mathcal{V}'$ are disjoint.


Our coding scheme is as follows:
\begin{enumerate}
\item For each connecting tree $(\mathcal{V}^\textnormal{T}_t, \mathcal{E}^\textnormal{T}_t)$, we transmit all $\{x_i \oplus x_j: (i,j) \in \mathcal{E}^\textnormal{T}_t\}$, i.e., we transmit the network-coded bits of the associated message pair for each edge. Note that we transmit $|\mathcal{V}^\textnormal{T}_t|-1$ bits.
\item For each message-connected leaf SCC $(\mathcal{V}^\textnormal{C}_c, \mathcal{A}^\textnormal{C}_c)$ (which is edge-connected by definition), we first obtain a spanning tree, denoted by $\mathcal{T}^\textnormal{ST}_c = ( \mathcal{V}^\textnormal{C}_c, \mathcal{E}^\textnormal{ST}_c)$, where $\mathcal{E}^\textnormal{ST}_c \subseteq \mathcal{E}$. We then transmit all $\{x_i \oplus x_j: (i,j) \in \mathcal{E}^\textnormal{ST}_c\}$. Note that we transmit $|\mathcal{V}^\textnormal{C}_c|-1$ bits.
\item For the rest of the non-leaf vertices, we transmit $\{ x_i: i \in \mathcal{V}'_\textnormal{out}\}$, i.e., we transmit the message bits uncoded.
\end{enumerate}
Each vertex in the connecting trees and the message-connected SCCs has at least one outgoing arc. Hence, the coding scheme generates an index code of length $V_\textnormal{out}(\mathcal{G}) - ( \connected + \tree)$.

We can easily verify that the index code can be transmitted, as each message pair to be XORed is associated with an edge, i.e., both the message bits belong to some sender.

Finally, we show that each receiver is able to obtain its required messages. Recall that each receiver $i$ needs to decode all messages in $\{ x_j: (j \rightarrow i) \in \mathcal{A}\}$. Now, each receiver $i$ must belong to one---and only one---of the following groups:
\begin{enumerate}
\item (Connecting tree) $i \in \mathcal{V}^\textnormal{T}_t$: Knowing $x_i$, receiver $i$ can decode all $\{x_j: j \in \mathcal{V}^\textnormal{T}_t\}$ from $\{x_j \oplus x_k: (j,k) \in \mathcal{E}^\textnormal{T}_t\}$ by traversing the tree (which is connected by definition).
It can also decode the messages $\{x_k: k \in \mathcal{V}'_\textnormal{out}\}$, sent uncoded.
Since all connecting trees and message-connected leaf SCCs have no outgoing arcs, each incoming arc to $i$ must be from either $\mathcal{V}^\textnormal{T}_t \setminus \{i\}$ or $\mathcal{V}'_\textnormal{out}$. So, receiver $i$ is able to decode all its required messages.
\item (Message-connected leaf SCC) $i \in \mathcal{V}^\textnormal{C}_c$: Using the same argument as that for the connecting trees, we can show that receiver $i$ can decode all its required messages.
\item (The remaining vertices) $i \in \mathcal{V}'$: Using the argument in point 1, all incoming arcs to vertex $i$ must come from $\mathcal{V}'_\textnormal{out} \setminus \{i\}$. Since we sent $\{x_j: j \in \mathcal{V}'_\textnormal{out}\}$ uncoded, receiver $i$ can decode all its required messages. $\hfill \blacksquare$

\end{enumerate}


\section{Discussions and Future Work}



We now show that the pairwise linear coding proposed in Sec.~\ref{section:achievable} can be suboptimal. Consider an index coding problem with $n=6$ messages/receivers, $\mathcal{A} = \{ (1 \leftrightarrow 2), (3 \leftrightarrow 4), (5 \leftrightarrow 6) \}$ where $(i \leftrightarrow j) \triangleq \{ (i \rightarrow j), (j \rightarrow i)\}$,
 and $S=4$ senders having the following messages $(x_1,x_3,x_5)$, $(x_3,x_5,x_2)$, $(x_5,x_2,x_4)$, and $(x_2,x_4,x_6)$ respectively.

In this example, there is no message-connected, message-disconnected, or degenerated semi leaf SCC in $\mathcal{G}$.
Running Algorithm~\ref{algorithm},  
we get the lower bound $\tilde{\ell}^* \geq 4$.


We can also show that the largest $\tree = 1$, i.e., the pairwise linear coding (Thm.~\ref{theorem:multi-sender-achievable}) can only achieve 5 bits. However, if each sender sends the XOR of its three message bits, the 4-bit lower bound is achievable.

This example illustrates that---in stark contrast to the single-sender case---the pairwise coding scheme described in Sec.~\ref{section:achievable} is not always optimal. It also shows a disadvantage of using $\mathcal{U}$: it cannot differentiate a sender having $(x_1,x_2,x_3)$ from three senders having $(x_1,x_2), (x_2,x_3), (x_1,x_3)$ respectively. For future work, we will investigate a more general coding scheme and a more informative graphical representation.

{\ }

\noindent {\hfill \scshape Appendix A \hfill}


Consider an index coding problem \textsf{IC1}, and denote its optimal codelength by $\tilde{\ell}^*(\textsf{IC1})$. Let $\mathcal{L}$ be the set of receivers whose messages $\{x_i: i \in \mathcal{L}\}$ are not required by any other receiver. 
Further consider the simplified index coding problem \textsf{IC2} where $x_i = \varnothing$ for all $i \in \mathcal{L}$. Denote its optimal codelength by $\tilde{\ell}^*(\textsf{IC2})$.


\begin{proposition} \label{proposition:simplified}
Any index code for \textsf{IC2} is an index code for \textsf{IC1}.
\end{proposition}

\begin{IEEEproof} 
  Any index code $\boldsymbol{c}$ for \textsf{IC2} can also be transmitted by the senders of \textsf{IC1}. 
 Since $\boldsymbol{c}$ satisfies the decoding requirement of \textsf{IC2}, it must also satisfy the those of \textsf{IC1}. 
\end{IEEEproof}

\begin{proposition} $\tilde{\ell}^*(\textsf{IC1}) = \tilde{\ell}^*(\textsf{IC2})$
\end{proposition}

\begin{IEEEproof} It follows from Proposition~\ref{proposition:simplified} that $\tilde{\ell}^*(\textsf{IC1}) \leq \tilde{\ell}^*(\textsf{IC2})$.
Now consider any optimal index code $\boldsymbol{c}^*$ for \textsf{IC1}. By definition it can contain $\{x_i: i \in \mathcal{L}\}$. Since $\boldsymbol{c}^*$ is an index code, all receivers in \textsf{IC1} can decode their required messages when $x_i = 1$ for all $i \in \mathcal{L}$ (these messages $x_i$ are not required by any receiver). So, $\boldsymbol{c}^*$ with all $x_i$ set to 1 is also an optimal index code. Denote this code by $\boldsymbol{c}'$. Now, since $\boldsymbol{c}'$ does not depend on the actual contents of $\{x_i: i \in \mathcal{L}\}$, $\boldsymbol{c}'$ can also be sent by the senders in \textsf{IC2}, and it also satisfies the decoding requirement of \textsf{IC2}. This means $\boldsymbol{c}'$ is an index code for \textsf{IC2}, and hence $\tilde{\ell}^*(\textsf{IC2}) \leq \tilde{\ell}^*(\textsf{IC1})$.
\end{IEEEproof}
{\ }

\noindent{\hfill \scshape Appendix B \hfill}

Recall that in the simplified graphs, if $i$ is a leaf vertex, then $x_i = \varnothing$, i.e., it has no prior message.
Consequently any receiver $p$, regardless of the prior knowledge $\mathcal{K}_p$ it possesses, is as good as the leaf vertex (receiver) $i$.
Hence $p$ must be able to decode all messages decodable by receiver $i$, and the result follows from the proof of Lemma~\ref{lemma:predecessor}.

{\ }

\noindent{\hfill \scshape Appendix C \hfill}

Given any digraph $\mathcal{G}$, we form a {\em supergraph} $\mathcal{G}_\textnormal{s}$ by replacing each (leaf or non-leaf) SCC with at least two vertices  by a (special) vertex referred to as a {\em supernode}. First, $\mathcal{G}_\textnormal{s}$ cannot contain any directed cycle. Otherwise, all supernodes and vertices in the cycle form an SCC, and it would have been collapsed into a supernode. Further, if $\mathcal{G}$ has no leaf SCC, meaning that $\mathcal{G}_\textnormal{s}$ has no leaf supernode, then every supernode and non-leaf vertex must have a path to a leaf vertex. This means $\mathcal{G}$ is grounded. $\hfill\blacksquare$


\begin{thebibliography}{1}
\providecommand{\url}[1]{#1}
\csname url@samestyle\endcsname
\providecommand{\newblock}{\relax}
\providecommand{\bibinfo}[2]{#2}
\providecommand{\BIBentrySTDinterwordspacing}{\spaceskip=0pt\relax}
\providecommand{\BIBentryALTinterwordstretchfactor}{4}
\providecommand{\BIBentryALTinterwordspacing}{\spaceskip=\fontdimen2\font plus
\BIBentryALTinterwordstretchfactor\fontdimen3\font minus
  \fontdimen4\font\relax}
\providecommand{\BIBforeignlanguage}[2]{{%
\expandafter\ifx\csname l@#1\endcsname\relax
\typeout{** WARNING: IEEEtran.bst: No hyphenation pattern has been}%
\typeout{** loaded for the language `#1'. Using the pattern for}%
\typeout{** the default language instead.}%
\else
\language=\csname l@#1\endcsname
\fi
#2}}
\providecommand{\BIBdecl}{\relax}
\BIBdecl

\bibitem{birkkol2006}
Y.~Birk and T.~Kol, ``Coding on demand by an informed source ({ISCOD}) for
  efficient broadcast of different supplemental data to caching clients,''
  \emph{IEEE Trans. Inf. Theory}, vol.~52, no.~6, pp. 2825--2830, June 2006.

\bibitem{elrouayheb10}
S.~{El Rouayheb}, A.~Sprintson, and C.~Georghiades, ``On the index coding
  problem and its relation to network coding and matroid theory,'' \emph{IEEE
  Trans. Inf. Theory}, vol.~56, no.~7, pp. 3187--3195, July 2010.

\bibitem{baryossefbirk11}
Z.~Bar-Yossef, Y.~Birk, T.~S. Jayram, and T.~Kol, ``Index coding with side
  information,'' \emph{IEEE Trans. Inf. Theory}, vol.~57, no.~3, pp.
  1479--1494, Mar. 2011.

\bibitem{dauskachekchee12}
S.~H. Dau, V.~Skachek, and Y.~M. Chee, ``On the security of index coding with
  side information,'' \emph{IEEE Trans. Inf. Theory}, vol.~58, no.~6, pp.
  3975--3988, June 2012.

\bibitem{ongho12}
L.~Ong and C.~K. Ho, ``Optimal index codes for a class of multicast networks
  with receiver side information,'' in \emph{Proc. IEEE Int. Conf. Commun.
  (ICC)}, Ottawa, Canada, June 2012, pp. 2223--2228.

\bibitem{oechteringschnurr08}
T.~J. Oechtering, C.~Schnurr, and H.~Boche, ``Broadcast capacity region of
  two-phase bidirectional relaying,'' \emph{IEEE Trans. Inf. Theory}, vol.~54,
  no.~1, pp. 454--458, Jan. 2008.

\bibitem{ongkellettjohnson12-it}
L.~Ong, C.~M. Kellett, and S.~J. Johnson, ``On the equal-rate capacity of the
  {AWGN} multiway relay channel,'' \emph{IEEE Trans. Inf. Theory}, vol.~58,
  no.~9, pp. 5761--5769, Sept. 2012.

\bibitem{lubertzkystav09}
E.~Lubetzky and U.~Stav, ``Nonlinear index coding outperforming the linear
  optimum,'' \emph{IEEE Trans. Inf. Theory}, vol.~55, no.~8, pp. 3544--3551,
  Aug. 2009.

\bibitem{Bang-JensenGutin}
J.~Bang-Jensen and G.~Gutin, \emph{Digraphs: Theory, Algorithms and
  Applications}.\hskip 1em plus 0.5em minus 0.4em\relax Springer Verlag, 2007.

\end{thebibliography}

\end{document}